# Explaining the Origin of the Anthropocene and Predicting Its Future

*By* Ron W. NIELSEN[†]

**Abstract.** Growth of the world population and the world economic growth were hyperbolic in the past 2,000,000 years. Recently, from around 1950, they started to be diverted to slower trajectories but they are still close to the historical hyperbolic trajectories. Regional growth of population and regional economic growth were also hyperbolic. Hyperbolic growth can be slow over a long time and fast over a short time but it is still the same, monotonically increasing, growth. It is incorrect to interpret slow growth as stagnation and fast growth as explosion, each controlled by different mechanisms of growth. Hyperbolic growth has to be interpreted as a whole and the same mechanism has to be applied to the slow and to the fast growth. The Anthropocene is characterised by the rapid growth of population, rapid economic growth and rapid consumption of natural resources. The origin of the Anthropocene can be explained as the natural consequence of hyperbolic growth. However, while its dramatic impacts became apparent only recently its starting point cannot be mathematically determined. Its beginning could be traced to the dawn of the existence of hominines. Its future is insecure because it is dictated by many critical trends shaping the future or our planet, but notably because the size of the world population is predicted to continue to increase at least until the end of the current century to a possibly unsustainable level and because the world economic growth follows now an unsustainable trajectory.

## 1. Introduction

Never before, in the long history of our planet, was there a single species that had such a profound impact on the environment as humans. We have a potential of converting this little speck of life in our Solar System to a hostile and inhabitable world but we also have a potential to survive and even prosper. Our origin can be traced to around 200,000 years ago or even earlier (Weaver, Roseman & Stringer, 2008), maybe even to between 2 and 3 million years ago, if we consider the origin of the genus *Homo*. However, our strong impact on the environment became apparent only recently, approximately two hundred years ago or even later.

Our impact is now so profound that a new geological era has been proposed and by now even is becoming widely accepted, the Anthropocene (Crutzen & Stoermer, 2000; Ehlers & Krafft, 2006; Steffen, Grinevald, Crutzen & McNeill, 2011; Zalasiewicz, Williams, Steffen & Crutzen 2010). This new epoch is described extensively in my book (Nielsen, 2006), which contains a comprehensive analysis of all critical trends shaping the future of our planet. I have divided them into seven groups, (1) the rapid growth of population generally described as the population explosion; (2) the diminishing land resources (3) the diminishing water resources, (4) the destruction of the atmosphere, (5) the approaching energy crisis, (6) social decline and (7) conflicts and the increasing killing power. This book is full of facts and figures, which can be used to understand the Anthropocene.

The primary driving force of the Anthropocene is the rapid growth of population. Consequently, in order to explain its origin, it is essential to understand the mechanism of growth of human population. We have to have clear understanding of this topic. Unfortunately, there are certain strongly misleading and scientifically unsupported misconceptions in this field of study (Nielsen, 2016a). The related issue is the economic growth, which turns out to follow similar time-dependent trajectories. This field is also affected by similar misconceptions (Nielsen, 2014, 2016a). They are all based on impressions and when closely analysed they are found to be contradicted by data. We cannot afford being guided by such misconceptions. The risk is too great.

[†] AKA Jan Nurzynski, Griffith University, Environmental Futures Research Institute, Gold Coast Campus, Qld, 4222, Australia.
☎. +61407201175
✉. ronwnielsen@gmail.com

Having described in my book the panorama of all critical trends shaping the future of our planet, I have devoted the past few years on the investigation of the growth of population and of the economic growth. I have formulated a general law of growth (Nielsen, 2016b) and I have used it to explain the mechanism of growth (Nielsen, 2016c). A compilation of my publications in these fields of study is now presented in a single document (Nielsen, 2017a). My new research is now focused on the investigation of the current economic growth and of the growth of population. I have formulated a mathematical method of analysis of growth trajectories and of forecasting (Nielsen, 2017b), which I use to analyse the current growth. The aim of this work is to understand the most likely trajectories of the future growth, to understand the warning signs and to understand what needs to be done to ensure a sustainable future.

Growth of human population in the past 2,000,000 years was slow (Nielsen, 2017c). The first billion of global population was reached around AD 1800 (Biraben, 1980; Durand, 1974; McEvedy & Jones, 1978; Thomlinson, 1975; United Nations, 1973, 1999). Thus, it took many thousands of years, indeed even millions of years, for the world population to increase to one billion but after reaching the first billion, the second billion was added in just only about 130 years (United Nations, 1999). The process of many thousands of years, or even millions of years, was suddenly compressed to just over 100 years. Consumption of natural resources and the stress on the environment increased rapidly. It might be surprising that we have survived this enormous stress.

If adding one billion in just 130 years sounds too fast, the next billion was added in just 29 years, the next in 15 years, the next in 13 years, and the next in 12 years, increasing the size of global population to 6 billion (US Census Bureau, 2017). The last billion, which boosted global population to 7 billion, was added in 13 years (US Census Bureau, 2017). We call it the slowing-down growth (because the last billion was added in 13 years rather than in 12 years or even in a shorter time) but obviously the slowing down process is still too slow.

When closely analysed, data show consistently that the growth of human population and the economic growth in the past 2,000,000 years were hyperbolic (Nielsen, 2016d, 2016e, 2016f, 2016g, 2016h, 2016i, 2016j, 2017c, 2017d). Hyperbolic growth started to be diverted gradually to slower, non-hyperbolic trajectories only recently, around 1950 (for the global growth of population and for the global economic growth) but the new trajectories are still close to the original hyperbolic trajectories. For the regional growth, the diversions occurred, in some cases, even earlier. The fast growth of population in recent years is the natural outcome of hyperbolic growth. Consequently, in order to explain the origin of the Anthropocene it is imperative to have clear understanding of hyperbolic growth. Indeed, the current prevailing misconceptions about the growth of population and about the economic growth are based on the incorrect understanding of hyperbolic growth.

Hyperbolic growth is slow over a long time and fast over a short time but it is still the same, monotonically increasing growth, which cannot be divided into two different components. Hyperbolic growth has to be interpreted as a whole. The same mechanism of growth has to be applied to the slow and to the fast growth.

It is incorrect to interpret the slow growth as stagnation and the fast growth as explosion, each governed by a different mechanism of growth. We can loosely describe the perceived fast growth as explosion as long as we understand that it is just a perceived feature, which did not occur at any specific time and that it is a feature, which is just *the natural continuation of hyperbolic growth*. It is controlled by precisely the same mechanism as the slow growth. The Anthropocene is the natural consequence of hyperbolic growth but, as we shall see, its beginning cannot be mathematically determined.

## 2. Hyperbolic distributions

In order to understand hyperbolic growth, it is convenient to compare it with the more familiar exponential growth. They are both presented in Figure 1.

Exponential growth is described by the following equation:

$$S(t) = ae^{kt}, \qquad (1)$$

where $S(t)$ is the size of the growing entity (in our case, the size of population or the size of the Gross Domestic Product (GDP), representing economic growth), $a$ is the normalisation constant, $k$ is the constant representing growth rate and $t$ is the time.

For the distribution shown in Figure 1, $a = 1.414 \times 10^{-1}$ and $k = 2.400 \times 10^{-3}$. The size $S(t)$ is in billions and the time is in years.

Hyperbolic growth is described by the following equation:

$$S(t) = \frac{1}{a - kt}. \tag{2}$$

Parameters $a$ and $k$ are positive constants but $k$ does not represent the growth rate. For the distribution shown in Figure 1, $a = 7.061 \times 10^{0}$ and $k = 3.398 \times 10^{-3}$.

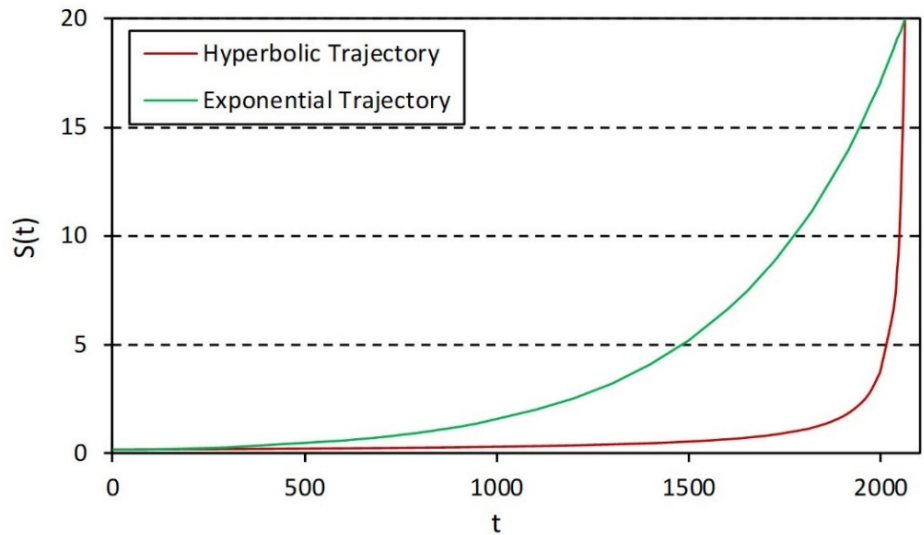

**Figure 1.** *Exponential and hyperbolic trajectories. Hyperbolic growth creates an illusion of an explosion at a certain time. The change of direction from the nearly horizontal to the nearly vertical growth is real but the time of the change cannot be mathematically determined. Hyperbolic growth increases monotonically.*

In the case of the exponential growth, there is no confusion. It is clear that its trajectory increases monotonically and that there is no explosion. However, hyperbolic growth might be puzzling and confusing. Such distributions create a serious problem in the demographic and economic research.

Hyperbolic growth is slow over a long time, so slow that it is approximately horizontal. Then, over a certain short time it appears to be changing its character and increases so fast that it becomes nearly vertical. It might be tempting to divide this distribution into two, or maybe even three components and assign to them distinctly different mechanisms of growth. However, such a division would be a serious mistake, the mistake which is unfortunately repeated in the demographic and economic research. Hyperbolic growth cannot be divided into different components and the best way to see it, is to display its reciprocal values:

$$\frac{1}{S(t)} = a - kt, \tag{3}$$

because in this form, hyperbolic growth is represented by a decreasing straight line, as shown in Figure 2. It is precisely the same distribution as shown in Figure 1 but now plotted differently. Properties of mathematical distributions do not change when they are plotted in different ways but certain features, which are not clear in one representation might become clear in another. It is, indeed, now clear that hyperbolic distributions cannot be divided into different components. This conclusion is so obvious that it is hard to understand how it was possible to create the Unified Growth Theory, which is firmly based on the assumption of the existence of three distinctly different regimes of growth (Galor, 2005, 2011). Mistakes can be made but mistakes can be also corrected and science has many examples of corrected mistakes.

Now it is also clear that it is impossible to identify the time or the small range of time on the hyperbolic distribution, which could be claimed as a transition from a slow to a fast growth. Which point on the straight line could be claimed as the time of an unusual acceleration?

The transition occurred over the whole range of hyperbolic growth. There was no unusual acceleration at any time because such an unusual acceleration would be reflected in a change of direction of the straight-line trajectory. The straight line remains undisturbed, which means that there was no unusual acceleration at any time.

The nearly vertical growth, which is usually described as explosion, is just the natural continuation of hyperbolic growth. It is represented by precisely the same straight line as the line corresponding to the slow growth, which means that the mechanism describing the slow growth must be the same as the mechanism describing the fast growth.

It is impossible to determine the time of transition from the slow to fast growth because there was no transition at any time but a gradual transition over the whole range of hyperbolic growth. If we accept that the fast growth of population identifies the Anthropocene, then it is now clear that its beginning cannot be mathematically determined.

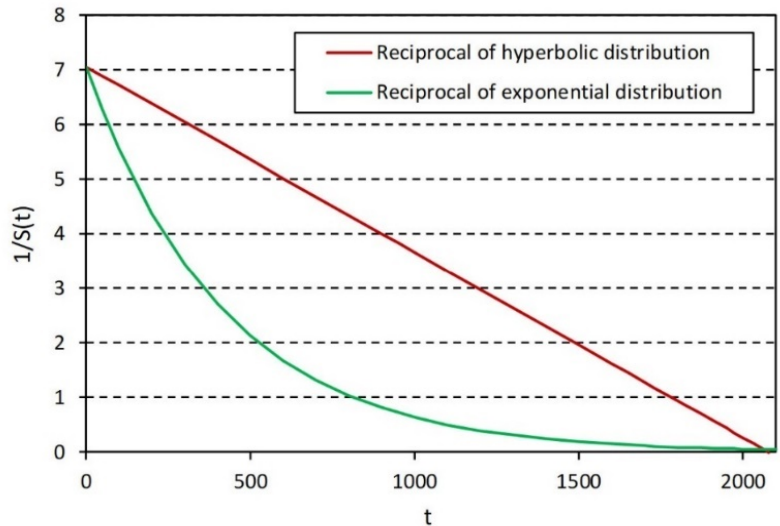

**Figure 2.** *Reciprocal values of hyperbolic growth show clearly that hyperbolic distribution cannot be divided into two or three different components because it makes no sense to divide a straight line into different components. The change of direction shown in Figure 1 did not occur at any specific time but over the whole range of hyperbolic growth. Reciprocal exponential distribution is also displayed. It is an exponentially decreasing distribution.*

Figure 2 shows also the reciprocal of the exponential growth, which is now represented by the exponentially decreasing distribution.

$$\frac{1}{S(t)} = \frac{1}{a} e^{-kt}. \tag{4}$$

We can take yet another approach to dispel the illusion of stagnation followed by a sudden explosion by using semilogarithmic set of coordinates, as shown in Figure 3.

In this form, exponential growth is represented by a straight line. Hyperbolic growth is now also clearly seen as a monotonically increasing distribution. It is again obvious that there is no unusual acceleration at any time. It is obvious that it would be futile to try to identify a place for a transition from a slow to a fast growth.

The illusion of a possible transition is only seen in Figure 1. Any attempt to determine a transition from a slow to a fast growth would be strongly subjective and mathematically unjustified. There is no mathematical criterium, which can be used to determine when hyperbolic growth changes from slow to fast. The change takes place monotonically over the whole range of hyperbolic distribution.

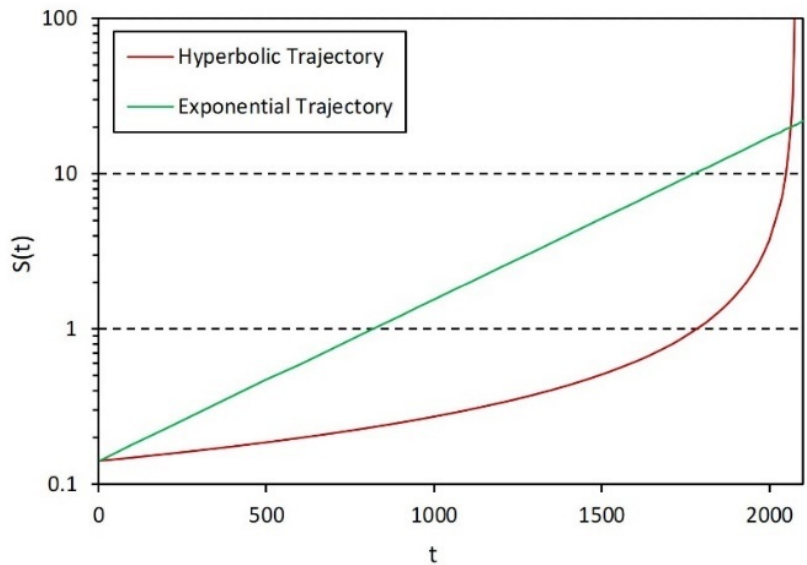

**Figure 3.** *Semilogarithmic representation of exponential and hyperbolic distributions. Hyperbolic distribution increases monotonically and the change of direction occurs along the whole range of growth.*

Even for the linear scales, such as used in Figure 1, the illusion of an unusual acceleration at a certain time depends on the compression of linear scales. Let us, for instance, magnify the part of the plot, which appears to be showing the unusual acceleration described commonly as explosion. Such magnification is shown in Figure 4, again using linear scales. Now we can see that the change from slow to fast growth occurs gradually without any unusual change of direction. Any attempt to determine the time of change or even the small range of time would be strongly subjective and mathematically unjustified.

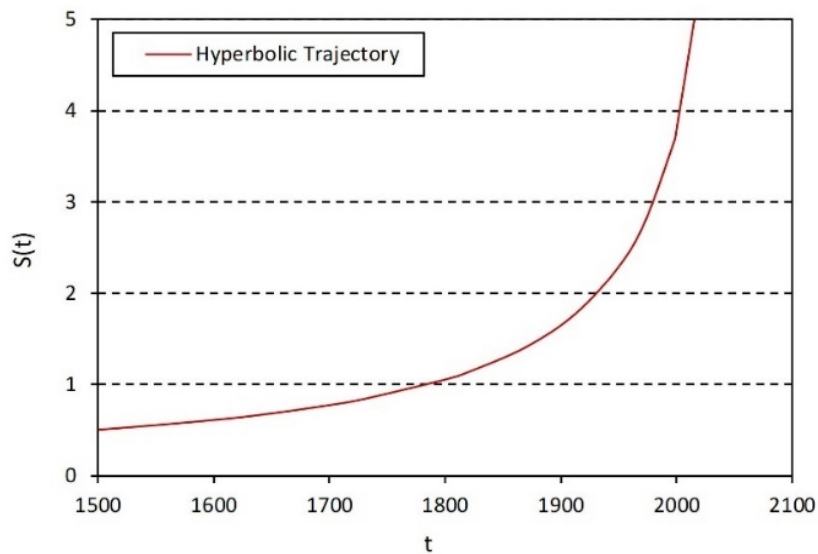

**Figure 4.** *Magnified part of the linear display used in Figure 1, focusing now on the section where there was an illusion of a change from slow to fast growth at a certain time. Even by using linear scales, the illusion of sudden explosion is replaced by the natural continuation of the same growth.*

## 3. Growth rate

In order to understand hyperbolic distributions, we have to understand also their growth rate. Again, we shall compare the growth rates of hyperbolic and exponential distributions.

Exponential growth is a solution of the following differential equation:

$$\frac{1}{S(t)}\frac{dS(t)}{dt} = k .\qquad(5)$$

The left-hand side of this equation is, by definition, the growth rate. For the exponential growth, the growth rate is constant. It does not matter how large is the size of the growing entity, the growth rate remains all the time the same, which means that the growth rate per element of the growing entity (in our case it would be per person or per dollar) decreases exponentially with the increasing size of the growing entity.

Hyperbolic growth is a solution of the following differential equation:

$$\frac{1}{S(t)}\frac{dS(t)}{dt} = kS(t) .\qquad(6)$$

It is just a small modification of the differential equation describing exponential growth but with profound consequences. For the hyperbolic distributions, the growth rate is not constant but directly proportional to the size of the growing entity. Now, the growth rate per element (per person or per dollar) of the growing entity is constant. It is a special kind of a self-propelling growth. Each element contributes equally to the growth process. The larger is the size of the growing entity, the larger is the combined force of growth and the faster is the growth. This is an important characteristic property, which identifies hyperbolic growth. This characteristic property can be used to explain the mechanism of the hyperbolic growth of human population. Likewise, the equivalent property that the growth rate is directly proportional to the size of the growing entity can be used to explain the mechanism of the hyperbolic economic growth.

We can now use parameters describing exponential and hyperbolic distributions shown in Figures 1-3 to calculate their corresponding growth rates. They are shown in Figure 5.

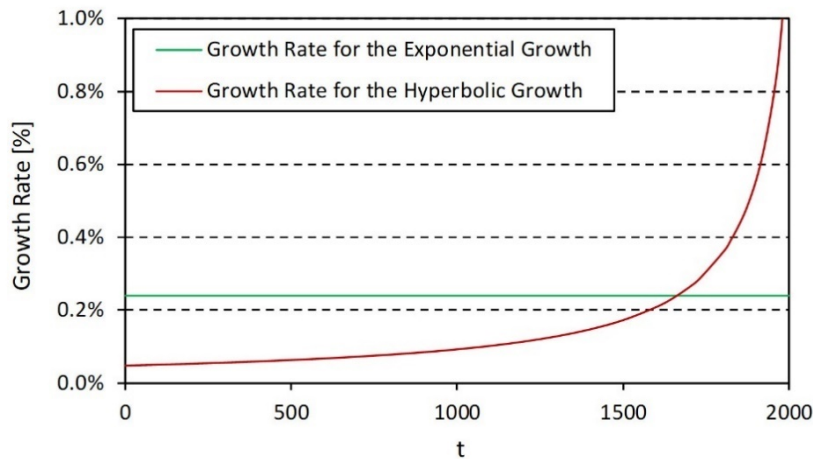

**Figure 5.** *Growth rates for the exponential and hyperbolic distributions displayed in Figure 1. Growth rate for the exponential distribution is constant, while the growth rate for the hyperbolic distribution increases monotonically without any signs of unusual acceleration at any time or over any small range of time. It is mathematically impossible to determine the time of transition from the slow to fast growth.*

We can see that while for the exponential distribution the growth rate is constant, for the hyperbolic distribution it increases monotonically with time. Indeed, for the hyperbolic growth its growth rate increases also hyperbolically. It is because the growth rate per element of the growing entity is constant that the combined growth rate increases relentlessly with time. It is mathematically impossible to determine the time of change from slow to fast growth. The change takes place monotonically over the whole range of hyperbolic growth.

We can also display the growth rate as the function of the size of the growing entity. Such a display is shown in Figure 6.

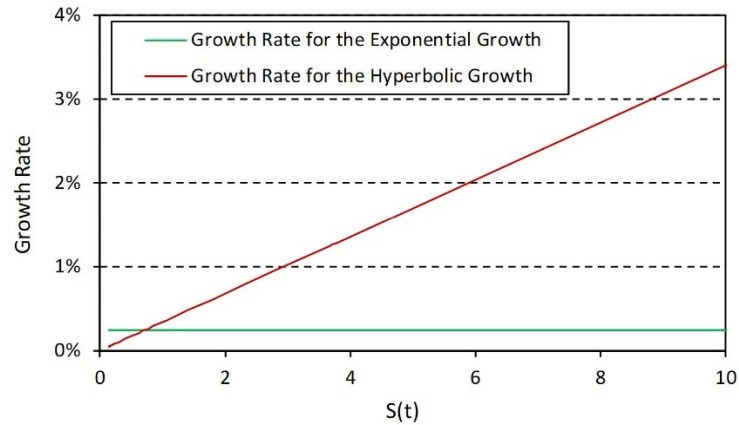

**Figure 6.** *Growth rates plotted as functions of the size of the growing entity. The growth rate for the exponential growth is constant. For the hyperbolic growth, it is directly proportional to the size of the growing entity. This plot demonstrates again that hyperbolic growth cannot be divided into two or three different regimens of growth, as erroneously assumed in the Unified Growth Theory (Galor, 2005, 2011). It also shows that there was no unusual acceleration in the growth rate at any time and thus that the transition from the slow to fast growth cannot be assigned to any time or to any small range of time. The transition was occurring over the whole range of hyperbolic distribution.*

It is because the growth rate per element is constant that the combined growth rate for the hyperbolic growth is directly proportional to the size of the growing entity, i.e. to $S(t)$, which in our case is the size of population or the size of the GDP. This display illustrates again that hyperbolic growth cannot be divided into two or three different components and a transition from slow to fast growth cannot be associated with any specific time or with any small range of time. The transition was gradual over the entire range of hyperbolic growth.

## 4. The doubling time

It is also useful to have clear understanding of the concept of the doubling time. It is important to understand that the concept of characterising growth in terms of the doubling time *applies only to the exponential growth. The doubling time should never be used to describe any other type of growth* because it is only for the exponential growth that the doubling time represents the characteristic feature of growth. Using the doubling time to characterise or describe non-exponential distributions is as useful as describing the cuttlefish as, for instance, a green animal. For any non-exponential distribution, the doubling time changes as we follow the trajectory of growth.

The doubling time for the exponential growth is given by the following formula:

$$T_2 = \frac{\ln 2}{k}. \tag{7}$$

This formula can be easily derived using eqn (1). If we express $k$ in percent, then this formula can be rewritten as

$$T_2 = \frac{69.3}{k'} \approx \frac{70}{k'}, \tag{8}$$

where $k'$ is now expressed in per cent. This is the so-called "rule of 70" which is routinely but erroneously applied to any type of growth but it *should be applied only to the exponential growth*. Applying it to any other type of growth is not only meaningless but also misleading because by applying it we assume that any other type of growth is exponential while it is not. If trying to characterise non-exponential growth by the doubling time is already bad enough, using the formula (7) or (8) is even worse. It simply makes no sense.

For the hyperbolic growth, the doubling time is not constant. It depends on time. Using eqn (2) we can show that the doubling time for the hyperbolic growth is given by

$$T_2 = \frac{1}{2}\left(\frac{a}{k} - t\right). \tag{9}$$

It is a mathematically correct formula but it has a limited application and consequently it should be used with care, if at all. For instance, for the hyperbolic distribution shown in Figure 1, at $t = 1000$, the doubling time is 540 years. It means that staring from precisely that time, hyperbolic growth would double in 540 years. However, if we started from $t = 1100$, then the doubling time would be 490 years. Only 100 years difference and the doubling time is already significantly different. If we started from $t = 2000$, the doubling time would be 40 years. Such calculations might be interesting, for whatever reason, but they could be of little use.

The doubling time is routinely used to project growth but if applied to any other growth than exponential, the projection is both meaningless and misleading. To project growth for any other type of growth, a simple but more sophisticated method of analysis should be used (Nielsen, 2017b). Prediction of growth cannot be based on the growth rate at a certain fixed time, as for the exponential growth, but on the investigation of the growth rate over a sufficiently long period of time

## 5. Hyperbolic growth during the AD era

Growth of population and of the GDP during the AD era are shown in Figures 7 and 8. Data (Maddison, 2010) are compared with hyperbolic distributions. More extensive analysis is presented in separate publications (Nielsen, 2016d, 2016e, 2016g, 2017c).

Parameters describing hyperbolic growth of population are: $a = 8.724 \times 10^0$ and $k = 4.267 \times 10^{-3}$. Parameters describing the growth of the GDP are: $a = 1.716 \times 10^{-2}$ and $k = 8.671 \times 10^{-6}$.

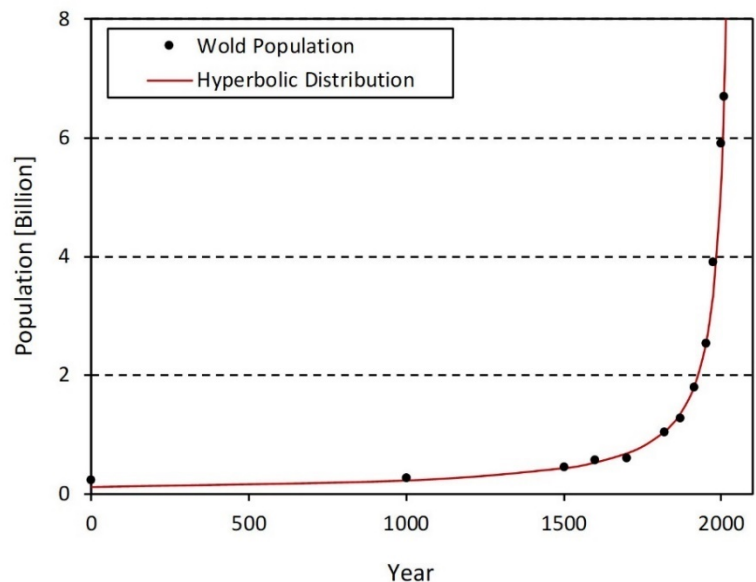

**Figure 7.** *Data describing growth of the world population (Maddison, 2010) are compared with hyperbolic distribution. Growth of population was not exponential, as incorrectly imagined by Malthus (1798) and as often claimed, but hyperbolic.*

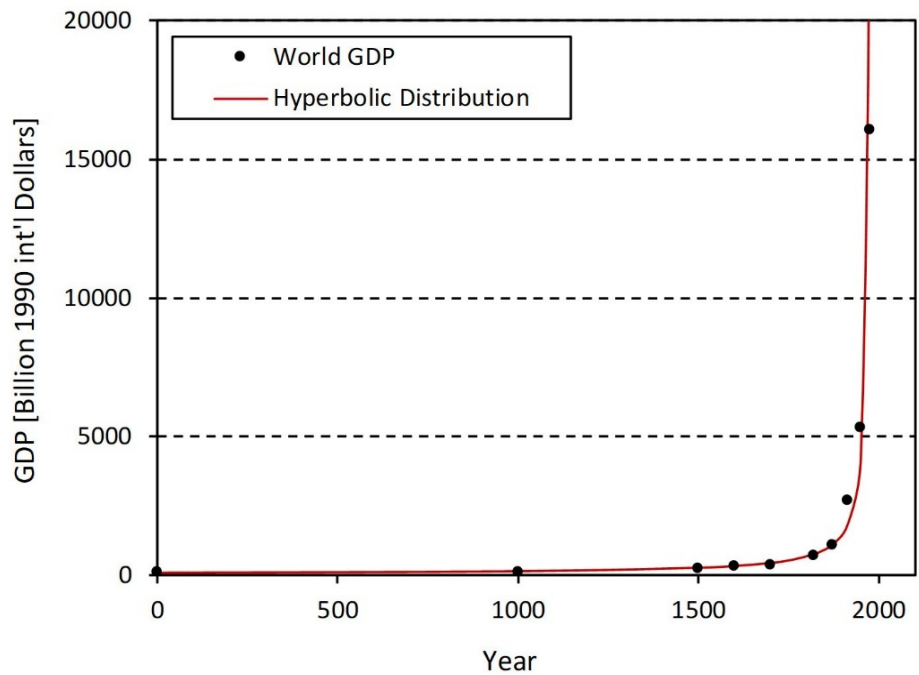

**Figure 8.** *Data describing growth of the world Gross Domestic Product (Maddison, 2010) are compared with hyperbolic distribution. Economic growth was hyperbolic.*

Hyperbolic growth of the world population was demonstrated as early as in 1960 (von Foerster, Mora & Amiot,1960). This important discovery is generally ignored. I have devoted the past few years of my life to remedy this unfortunate neglect and to help to understand the growth of population and economic growth, the two crucial processes propelling the Anthropocene and shaping our future.

1. I have introduced a simple method of analysis of hyperbolic distributions (Nielsen, 2014, 2017e).
2. I have demonstrated that hyperbolic description of the growth of population applies not only to the growth of the world population, as studied by von Foerster, Mora and Amiot (1960), but also to the growth of regional populations (Nielsen, 2016g).
3. I have demonstrated that hyperbolic description of growth applies also to the economic growth, global and regional (Nielsen, 2016d).
4. I have extended the analysis of the growth of the world population to the BC era (over the past 12,000 years) and demonstrated that hyperbolic growth applies not only to the AD era, as pointed out by von Foerster, Mora and Amiot (1960), but also to the BC era (Nielsen, 2016e).
5. I have extended the analysis of the growth of population and of the economic growth over a longer time (i.e. over the past 2,000,000 years) and confirmed the earlier observation of Deevey (1960) that the growth of population was in three major stages, but I have demonstrated that these stages were hyperbolic (Nielsen, 2017c).
6. I have explained the puzzling features of income per capita (GDP/cap) distributions by showing that they represent nothing more than mathematical properties of dividing two hyperbolic distributions (Nielsen, 2017d).
7. I have examined Galor's mysteries of growth (Galor, 2005, 2011) and explained them by showing that they are based on the incorrect understanding of hyperbolic distributions (Nielsen, 2016h, 2016j, 2017e) and on his habitual distorted representations of data.
8. I have demonstrated that the Unified Growth Theory describing economic growth and the Demographic Transition Theory describing the growth of population are based on the incorrect understanding of hyperbolic distributions and are contradicted by data (Nielsen, 2014, 2016d, 2016f, 2016g, 2016l, 2016m).

9. I have analysed the effects of the Malthusian positive checks and demonstrated that they have a dichotomous impact on the growth of population (Nielsen, 2016k) and thus, I have confirmed the earlier observation of Malthus (1798) about his positive checks and supported them by data, observation which is also unfortunately generally ignored.
10. I have demonstrated that, with only one exception demonstrated in a weak and short-lasting impact, demographic catastrophes did not shape the growth of population (Nielsen, 2016e, 2017f).
11. I have repeatedly demonstrated that Industrial Revolution had no impact on shaping growth trajectories, even in Western Europe and even in the United Kingdom (Nielsen, 2014, 2016d, 2016f, 2016g, 2016i, 2016m, 2016n, 2017e,).
12. I have formulated a general law of growth (Nielsen, 2016b) and used it to explain the mechanism of hyperbolic growth of human population and of economic growth (Nielsen, 2016c).

If hyperbolic distributions are confusing, they a significantly simpler than that the distributions describing income per capita represented by the GDP/cap. This issue was discussed in separate publications (Nielsen, 2016h, 2017d). Income per capita distributions (empirical and theoretical) for the world economic growth during the AD era are shown in Figure 9. They were obtained by dividing the relevant distributions shown in Figures 7 and 8.

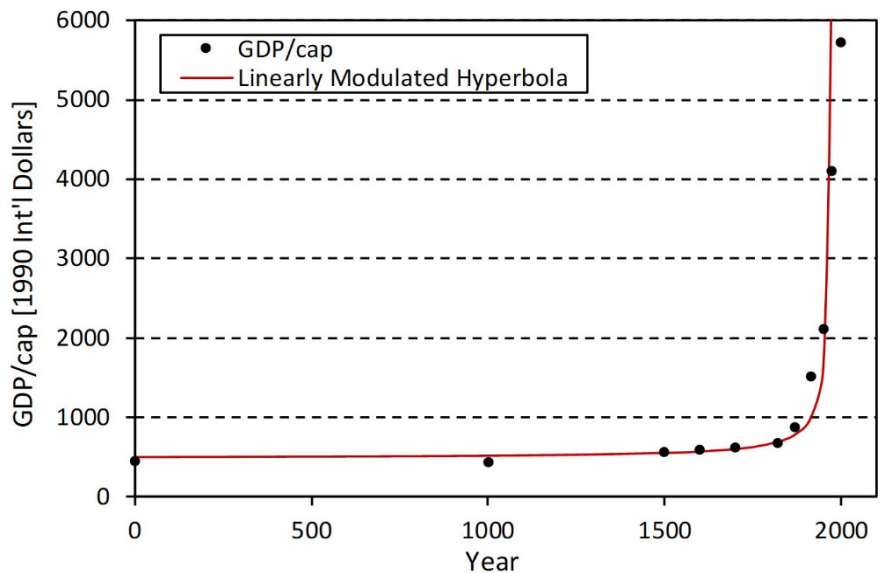

**Figure 9.** *Income per capita distributions (GDP/cap) obtained by dividing distributions shown in Figures 7 and 8. The puzzling features of growth (the approximately constant value over a long time followed by a rapid increase over a short time) represent nothing more than just the mathematical properties of dividing two hyperbolic trajectories (Nielsen, 2017d). The best fit to the data is simply a linearly-modulated hyperbolic distribution. Distributions shown in Figures 7 and 8 increase monotonically and consequently the distributions describing the growth of income per capita also increase monotonically, which can be easily confirmed by the analysis of the growth rate and of the gradient of the GDP/cap growth (Nielsen, 2017d). There was no stagnation and no explosion at any time.*

Distributions describing income per capita are nothing more than just linearly-modulated hyperbolic distributions. They also increase monotonically and there is no transition from a slow to fast growth at any time or over any small range of time, even though the linear display creates a strong illusion of such a transition.

There is nothing profoundly mysterious about the income per capita distributions. Their puzzling properties are nothing more than just mathematical properties of dividing two hyperbolic distributions. It is as simple as that.

The shape of the ratio of two hyperbolic distributions depends on the relative position of their singularities. If the singularity for the population distribution were earlier than the singularity for the GDP distribution, then the GDP/cap distribution would not be increasing to infinity but it would be decreasing to zero. For the distributions displayed in Figures 7 and 8, singularities are: $t_s = 1979$ for the GDP and $t_s = 2045$ for the growth of population. The distribution representing the GDP/cap ratio escapes rapidly to infinity in 1979. If the singularities were reversed, the distribution representing the GDP/cap would decrease rapidly to zero in 1979.

The mechanism of growth of the GDP/cap ratio can be easily explained. It is the same mechanism which describes the growth of the GDP and the growth of population, and the mechanism is exceptionally simple (Nielsen, 2016c; see section 7), which is hardly surprising because mathematical formula describing hyperbolic growth is also simple.

The fast-increasing income per capita is undeniably real but it did not start at any specific time and neither was it caused by any mysterious force. It is the natural consequence of the action of precisely the same forces that prompt hyperbolic growth of population and hyperbolic economic growth, forces that change monotonically and produce monotonically increasing trajectories. There is no need for introducing or for imagining any other additional force acting at any specific time to cause this rapid increase. There is no triggering mechanism, no ignition and no sudden explosion.

The rapid increase is the feature, which represents the natural consequence of the monotonically increasing hyperbolic distributions, and the characteristic shape of the GDP/cap distribution is nothing more than just the mathematical property of dividing two, monotonically increasing, hyperbolic distributions. We can take any two hyperbolic distributions described by the eqn (2) and divide them. As long as the singularity of the numerator is earlier than the singularity of the denominator, we shall get the characteristic features of the GDP/cap distribution. The mystery of the GDP/cap distributions is solved and the solution is simple.

## 6. Hyperbolic growth in the past 2,000,000 years

Growth of population in the past 2,000,000 years is shown in Figure 10 (Nielsen, 2017c). Similar trajectory is for the economic growth.

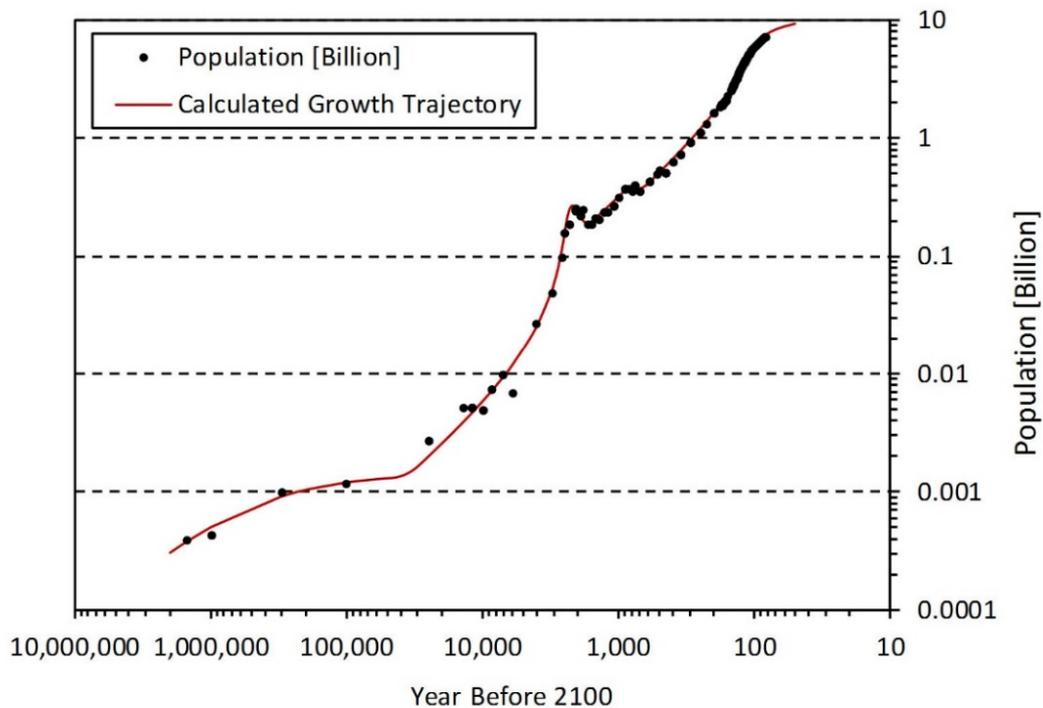

**Figure 10.** *Growth of human population in the past 2,000,000 years.*

This study confirms the earlier observation (Deevey, 1960) that the growth of population over such a long time was in three stages. However, while Deevey imagined that each state was leading to an equilibrium, my analysis shows that each stage was hyperbolic. This can be seen explicitly in Figure

11. The first stage does not look like hyperbolic in this graph but it is hyperbolic (see the explanation in my publication, Nielsen, 2017c).

It is remarkable that hyperbolic growth was so stable over such a long time. There were only two major transitions in the past 2,000,000 years. The first transition was between 46,000 BC and 27,000 BC and it was a transition from a slow to a faster hyperbolic growth. The second transition was between 425 BC and AD 510. It was a transition from a fast to a significantly slower hyperbolic trajectory.

There was also a minor disturbance during the AD era, which occurred between AD 1195 and 1470. This transition represents one and only example of impacts of demographic catastrophes (Nielsen, 2016e, 2017c, 2017f). A convergence of five major demographic catastrophes were needed to cause a minor and short-lasting disturbance in the growth of population, the disturbance, which was soon compensated by a faster growth, reflecting the stimulating effects of Malthusian positive checks (Malthus, 1798; Nielsen, 2016k). This transition changed the earlier hyperbolic trajectory to a slightly faster trajectory. The overall characteristics of the two trajectories are so similar that they can be replaced by a single trajectory fitting the data reasonably well (Nielsen, 2016e).

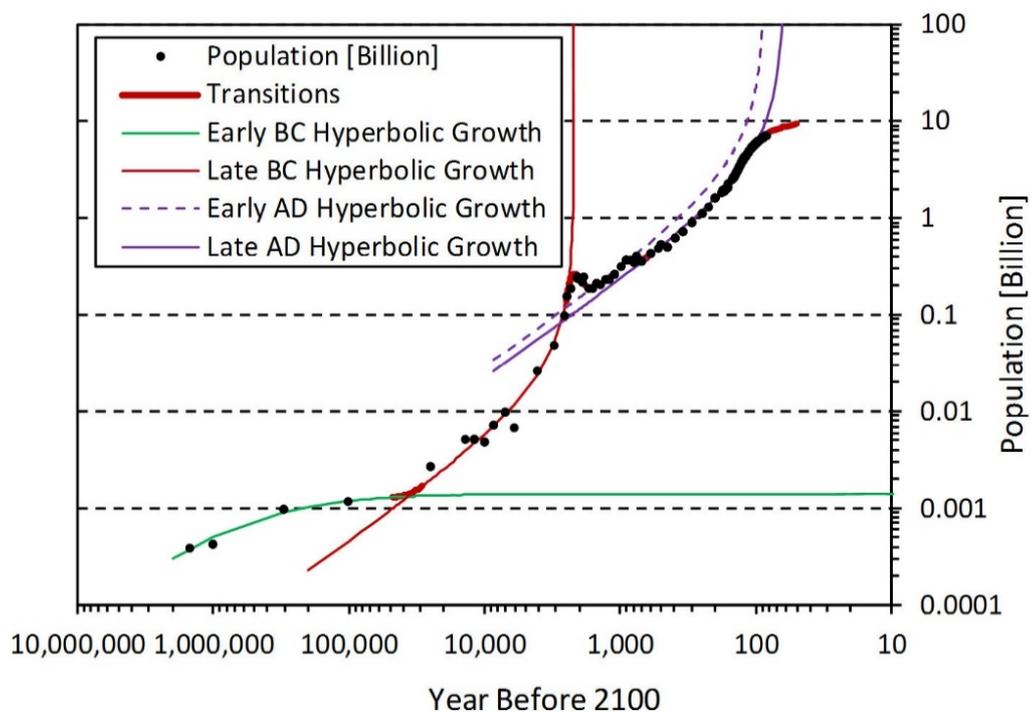

**Figure 11.** *Three major stages of growth of the world population in the past 2,000,000 years: (1) between 2,000,000 BC and 27,000 BC, (2) between 27,000 BC and AD 510 and (3) between AD 510 BC and the present time. Each stage ends with a transition to a new stage. The last stage experienced a minor distortion between around AD 1195 and 1470. This distortion caused a small shift in the hyperbolic growth. With just these three interruptions, growth of population in the past 2,000,000 years was hyperbolic leading inevitably to the Anthropocene. This proposed new epoch is the natural consequence of hyperbolic growth but its onset cannot be mathematically determined.*

Currently, beginning in around 1950, there is now a new transition to a yet unknown trajectory. This transition appears to be shaped by the increasing application of the Malthusian preventative checks but is moderated by the stimulating effects of the Malthusian positive checks in poor countries (Nielsen, 2016k).

Growth of population and economic growth in the past 2,000,000 was consistently hyperbolic. It was interrupted a few times but it soon resumed its hyperbolic growth. Relentlessly and persistently, it was increasing the size of the world population.

The characteristic feature of hyperbolic growth is that it contains singularity, a fixed time when the size of the growing entity increases to infinity. Close to the singularity, the size increases so fast that it can become uncontrollable. Nothing can increase to infinity and consequently, such a fast growth has to be terminated either catastrophically or by gradually diverting it to a different trajectory.

The late BC hyperbolic growth was so fast that, if continued, would have increased to infinity in 104 BC. Fortunately, it started to be diverted to a new trajectory in around 425 BC. The singularity was bypassed by a large margin of around 321 years. At the time of the commencement of this major transition, the size of the wold population was small, only around 140 million.

The point of singularity for the later AD hyperbolic growth is in 2037 but it was bypassed by 87 years when the growth started to be diverted to a new trajectory around 1950. However, this diversion started with a minor boosting, which brought the growth of population closer to the point of singularity. Furthermore, the size of the world population in 1950 was around 2.5 billion, much larger than the size of population in 425 BC. We are now in a far worse position with controlling the growth of population and with diverting it to a new and safe trajectory.

The Anthropocene is characterised by the exceptionally intensive impact of humans on the environment, the impact, which appears to have emerged suddenly in just the matter of only a few hundred years. However, such an apparent sudden appearance is the illusion created by hyperbolic growth. What we observe as a sudden phenomenon is in fact the natural consequence of hyperbolic growth. It appears to be sudden but it is not. The origin of the Anthropocene and its apparent sudden appearance, can be explained as the natural and inevitable consequence of hyperbolic growth but its beginning cannot be mathematically determined. This apparent sudden phenomenon had no beginning in recent time. If we want to trace its beginning we have to move to the dawn of hominine existence.

## 7. The mechanism of hyperbolic growth

Every growth can be, and usually is, described using growth rate. This is the quantity, which characterises growth. The larger is the growth rate, the faster is the growth.

Every growth is prompted by some kind of a force. We can assume, and indeed it appears to be obvious, that the driving force of growth is reflected or encoded in the growth rate. The stronger is the force of growth the larger is the growth rate and the faster is the growth. Furthermore, the force of growth determines and describes the mechanism of growth.

We might imagine a variety of correlations between the force of growth and the growth rate but the simplest correlation is represented by a force directly proportional to the growth rate:

$$F(t) = \kappa R(t), \qquad (10)$$

where $\kappa$ can be described as the resistance to growth and $R(t)$ is the growth rate,

$$R(t) = \frac{1}{S(t)} \frac{dS(t)}{dt}. \qquad (11)$$

In science, simple descriptions are tried first because natural phenomena can be usually described using simple principles. Complicated descriptions are avoided. They are introduced only if simple descriptions are inapplicable and they often suggest that we are on a wrong track and that we should look for an alternative simple description.

Equation (10) represents the general law of growth (Nielsen, 2016b). This law can be used to formulate a variety of models of growth. Possibilities are virtually unlimited.

The advantage of using this law of growth is that it links growth trajectories with the force of growth, which represents the mechanism of growth. Calculated trajectories might be represented by complicated mathematical formulae but by using this law of growth they can be usually based on simple assumptions about the force of growth or at least by assumptions that can be easily interpreted and easily related to the real life.

Thus, for instance, logistic model of growth is based on the assumption that the force of growth decreases linearly with the size of the growing entity. However, we might imagine a variety of other shapes for the trajectories describing the driving force. This force might, for instance, remain approximately constant over a certain range of sizes and then it might start to decrease gradually, faster or slower. Another possibility is that again the force of growth might not decrease linearly but approximately exponentially with the size of growing entity. Here, again, even if we assume a fixed starting point for the force of growth and a fixed, asymptotically approachable limit to growth, we can have various representations of the force of growth, each producing a different model of growth but all of them belonging to the same family of models, which resemble the logistic model of growth but are represented by different mathematical formulae. Our task then is not to explain the complicated mathematical descriptions of growth trajectories but the significantly simpler and maybe even intuitively understood descriptions of the driving force.

Often, even if we use simple representations of the driving force, mathematical description could be so complicated that the relevant differential equations have to be solved numerically. However, the starting assumptions can be usually simple. Again, in order to explain the mechanism of growth we do not have to explain the mathematically complicated descriptions of growth trajectories but the simple descriptions of the driving force. This approach simplifies considerably the study of the mechanism of growth.

The fundamental force of growth of human population is obviously the *biologically-controlled force of procreation* expressed as the difference between *the biologically-controlled sex drive*, which is ultimately related to birth rate and *the biologically-controlled process of aging and dying*, reflected in death rate. It would be unrealistic to expect that we should provide microscopic mathematical description of these biologically controlled forces. Such a task would be impossible and even if attempted it would quickly lead to some extremely complicated and incomprehensible formulations. Maybe by being so complicated they would be sufficiently impressive to be readily publishable in peer-reviewed journals but we would learn nothing useful from them. However, the net effects of these biologically controlled forces can be mathematically modelled in a simple, comprehensive and convincing way.

We can imagine many other forces controlling or prompting the growth of human population but by following the fundamental principle of parsimony in scientific investigations, we should consider first only the force of procreation and add to it other forces, if necessary. This force is essential and it cannot be replaced by other force of forces. We can add other forces to this force but we cannot replace it by any other force.

The simplest way to model the net effect of the biologically prompted force of procreation is to assume that on average it is *constant per person*:

$$\frac{F(t)}{S(t)} = c, \qquad (12)$$

where $c$ is a positive constant.

If we now use this definition of the driving force and insert it into the general law of growth given by the eqn (10), we shall get the eqn (6), which describes hyperbolic growth. In this equation, $k = c/\kappa$.

Thus, by assuming the simplest possible force of growth, we have now *derived* the hyperbolic growth equation. No other force is needed. *Hyperbolic growth of population represents the simplest possible mechanism of growth*, the unconstrained and spontaneous growth prompted by the, on average, constant per capita biologically controlled force of procreation.

If other forces make a significant contribution to the growth process, the growth is no longer hyperbolic. These rare exceptions occurred only three times in the past. The first time during the transition from a slow to a much faster hyperbolic trajectory between 46,000 BC and 27,000 BC, the second time between 425 BC and AD 510 and the third time between AD 1195 and 1470. Currently, starting from around 1950, the growth of population is also no longer hyperbolic, even though it is still following closely the original hyperbolic trajectory. We are in the process of a new transition to a yet unknown growth. It remains to be seen whether this will be short or long transition.

Analysis of data does not allow for the determination of parameters $c$ and $\kappa$ but only of their ratio, $c/\kappa$. However, for convenience, in the interpretation of the empirically determined parameter $k$, we might assume that the parameter $c$ remains the same all the time and that only the resistance to growth is changing. Hyperbolic growth between around 2,000,000 BC and 46,000 BC was characterised by an exceptionally large resistance to growth. During the transition between 46,000 BC and 27,000 BC, resistance to growth was undergoing a major adjustment to a new value. From around 27,000 BC, the resistance to growth was exceptionally low and the hyperbolic growth was fast (as measured by the parameter $k$) until around 425 BC, which marked the onset of a new adjustment in the resistance to growth. The new adjustment continued until around AD 510. From around that year, growth of human population settled along a new but slower hyperbolic trajectory (again as measured by the parameter $k$). During the minor disturbance between AD 1195 and 1470 the resistance to growth increased but the new hyperbolic trajectory from around AD 1470 was characterised by a slightly lower resistance to growth than the trajectory before AD 1195, reflecting the regenerating effects of the Malthusian positive checks (Malthus, 1798; Nielsen, 2016k). The two hyperbolic trajectories, before AD 1195 and after AD 1470 are virtually identical. They are only slightly shifted in time and they can be replaced by a single trajectory (Nielsen, 2016e). However, the two slightly shifted hyperbolic trajectories give a better description of data.

Equation describing hyperbolic growth [see eqn (2)] is exceptionally simple and it is therefore hardly surprising that the mechanism of hyperbolic growth is also simple. Explanation of the mechanism of the historical economic growth is also simple.

It is well known that wealth generates wealth. Again, any attempt to describe mathematically the microscopic interactions of all market forces would lead to incomprehensible, unconvincing and unserviceable mathematical formulations but we can model mathematically their net effect. We might consider a variety of factors contributing to the generation of wealth but the simplest possible mechanism is based on the assumption that, on average, the generated wealth is directly proportional to the already existing wealth. Under this simplest assumption, the force of the unconstrained economic growth is given by

$$F(t) = \rho W(t), \tag{13}$$

where $\rho$ is a constant and $W(t)$ is the existing wealth.

If we insert this force into the eqn (10), we shall get

$$\rho W(t) = \kappa R(t), \tag{14}$$

where

$$R(t) = \frac{1}{W(t)} \frac{dW(t)}{dt}, \tag{15}$$

which now leads us to the equation

$$\frac{1}{W(t)} \frac{dW(t)}{dt} = kW(t), \tag{16}$$

where $k = \rho/\kappa$.

This is again the equation describing hyperbolic growth. Thus, by assuming the simplest possible mechanism of economic growth we have *derived* the differential equation describing hyperbolic growth. Again, hyperbolic growth describes the simplest process of growth. More complicated descriptions of the force of growth are unnecessary.

## 8. The future of the Anthropocene

The future of the Anthropocene is dictated by the current convergence of critical trends shaping our future (Nielsen, 2006) but most notably by the growth of population and by the economic growth. We shall now examine what we can expect in these two areas. Calculations are based on my method of analysis of growth rates (Nielsen, 2017b).

*8.1. The growth of population*

The top part of Figure 12 shows the growth rate for the growth of the world population calculated directly from the population data (US Census Bureau, 2017). From around 1963, the growth rate was systematically decreasing. We can use these data to project growth.

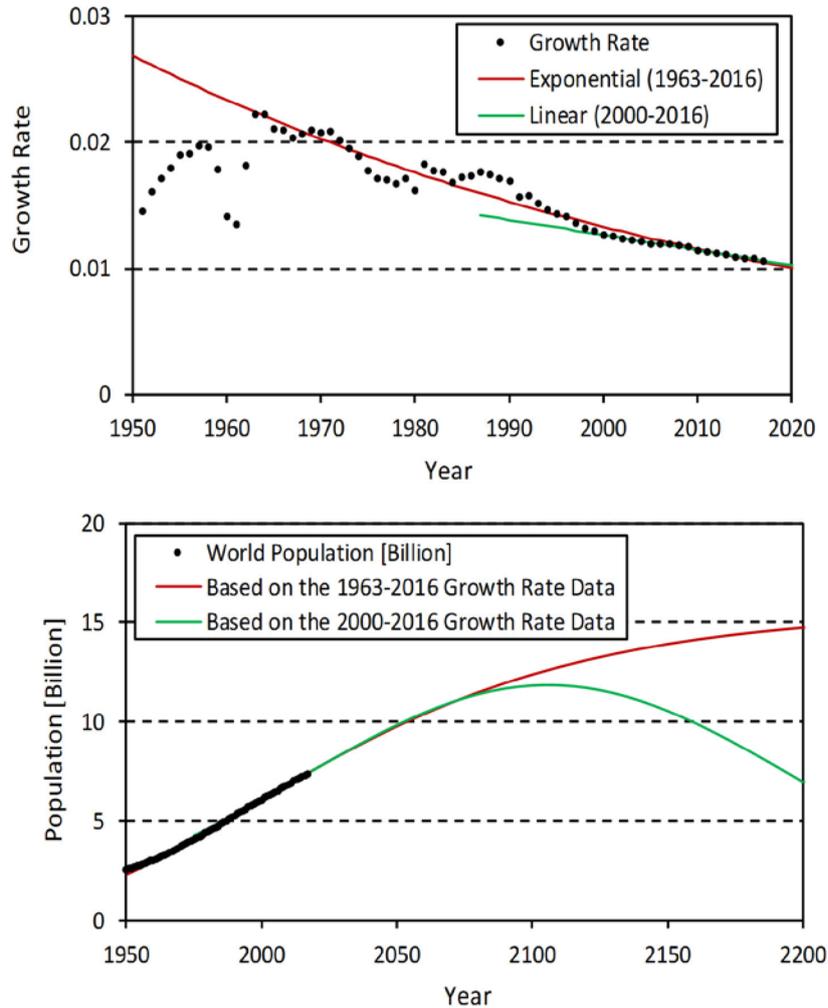

**Figure 12.** *Forecasting of the growth of the world population. Two representations (exponential and linear) of the growth rate, calculated using the US Census Bureau (2017) data, are used to generate growth trajectories for the growth of the world population (Nielsen, 2017b, 2017e). These calculations are in good agreement with projections of the United Nation (2015). However, the UN publication gives no information about the growth of the population in the 22nd century. It is important to notice that while the growth rate continues to decrease, the growth of the world population continues to increase and is not yet levelling off. It is projected to increase at least till the end of the current century.*

We seem to have two obvious options for the mathematical description of the growth rate: (1) to use the wide range of growth rate data between 1963 and 2016, which can be well described by the exponential function or (2) to assume that from around 2000 growth rate is now settling along a linearly decreasing trajectory. The projection of growth of the world population based on fitting exponential distribution to the growth rate could be regarded as more reliable because it is based on a wide range of data but it is still possible that the growth rate will now follow a linearly decreasing trajectory.

Results of calculations are shown in the lower part of Figure 12. If the growth rate is represented by the exponential function,

$$\frac{1}{S}\frac{dS}{dt} = ae^{bt}, \tag{17}$$

then the solution of the eqn (17) is (Nielsen, 2017b):

$$S = C\exp\left(\frac{a}{b}e^{bt}\right). \tag{18}$$

For the decreasing exponential function representing the growth rate (see the top section of Figure 12), the parameter $b$ is negative and the exponential function in the eqn (17) decreases asymptotically to zero. Consequently, $S$ increases asymptotically to the normalization constant $C$. The eqn (18) could be described as a pseudo-logistic trajectory. For the world population, parameters of this trajectory are: $a = 2.179 \times 10^{10}$ and $b = -1.406 \times 10^{-2}$, and its asymptotic value is 15.6 billion. The projected population in 2200 is 14.7 billion and increasing.

If we assume that the growth rate decreases linearly with time, i.e. if

$$\frac{1}{S}\frac{dS}{dt} = a + bt, \tag{19}$$

then the solution of this equation is (Nielsen, 2017b):

$$S(t) = C\exp\left[at + 0.5bt^2\right]. \tag{20}$$

Its parameters are: $a = 2.520 \times 10^{-1}$ and $b = -1.197 \times 10^{-4}$. It reaches a maximum of 11.8 billion in 2105.

Calculations shown in Figure 12 are in good agreement with projections by the United Nations (2015). According to this source "The world population is projected to increase by more than one billion people within the next 15 years, reaching 8.5 billion in 2030, and to increase further to 9.7 billion in 2050 and 11.2 billion by 2100" (United Nations, 2015, p. 2). My prognosis is 8.4 billion in 2030, 9.8 billion in 2050 and 11.8 billion in 2100 for the trajectory leading to the localized maximum. If the growth of the world population is going to follow the trajectory leading to the asymptotic maximum, then it will also reach 8.4 billion in 2030 and 9.8 billion in 2050 but only a slightly larger size of 12.4 billion in 2100. The difference between predicted values in 2100 is so small that we shall not know until well into the next century whether we are likely to reach a localized maximum of around 12 billion or to have the population continually increasing to the asymptotic size of around 16 billion, if such a large size can be supported by the accessible resources.

Summary of all these predictions is presented in Table 1. The UN projection gives no information about the expected size of population in the 22nd century. For the 21st century, the agreement between these two independent predictions is remarkably good.

**Table 1.** *Predicted growth of the world population*

| Source | 2030 | 2050 | 2100 | $S_{max}$ | $S_a$ |
|---|---|---|---|---|---|
| UN | 8.5 | 9.7 | 11.2 | NI | NI |
| CA | 8.4 | 9.8 | 11.8 | 11.9 | NA |
| CA | 8.4 | 9.8 | 12.4 | NA | 15.6 |

UN – United Nations, 2015; CA – current analysis (Nielsen, 2017b); NI – no information; NA – not applicable; $S_{max}$ – maximum value; $S_a$ – asymptotic value

The future of the Anthropocene is uncertain. Using the most optimistic prediction, the maximum size of the world population will be around 12 billion by the end of the current century or at the beginning of the next century, i.e. around twice as high as around 2000. Shall we be able to support such a large number of people? If not, we might expect a serious crisis.

However, there is also a possibility that the world population will not reach a maximum but will continue to increase to its asymptotic value of around 16 billion. By the end of the next century it might be close to 15 billion, i.e. about twice as large as the current (in 2017) world population. Shall we be able to support such a continuing growth?

The best option, if we had an option, would be to try to slow down the growth of population even more than now experienced, but we can hardly expect that such a global undertaking will be ever attempted, or even if undertaken that it would be ever successful. It is hard to control the growth of a large size of the increasing population, and an excellent example is China. They made a determined effort to control the growth of their population and they managed to reduce their growth rate to around 0.5% from a maximum of 1.6% in 1988 (World Bank, 2017). The growth rate remained constant at around 0.5% for the past 10 years, but recently it started showing signs of a gradual increase. If the growth rate is going to remain constant, the growth of population in China will be exponential and it will never level off. If the growth of population in China is going to continue at the approximately 0.5%, its size will increase to around 2.2 billion in 2100 and to 3.6 billion in 2200. Compared with the size of the population in 2000, it will be about 65% higher in 2100 and 170% higher in 2200. By around 2140, the population in China will be approximately twice as high as in 2000.

If we wanted to control the growth of the world population, we would have to undertake a massive and consolidated effort. There are no signs that we shall ever do it. The evidence-based best option would be to improve the living conditions in poor countries because Malthusian positive checks stimulate growth (Malthus, 1798; Nielsen, 2016k), but there are no indications that this possible solution will be ever attempted. Furthermore, improving their standard of living can be only achieved by improving their economic status, which would have to be now at the cost of the economic sacrifice of richer countries, because the current global economic growth follows already an unsustainable trajectory (Nielsen, 2015).

It is also important to notice that while the growth rate is decreasing the size of the world population is increasing. It is not yet levelling off. The size of the population will start to decrease if its growth rates is going to becomes negative. It will start to level off only if its growth rate will be approaching asymptotically the zero value. If the growth rate is going to decrease asymptotically to a positive constant value, the growth of the world population will be exponential.

## 8.2. Economic growth

Growth rate describing the growth of the world Gross Domestic Product (GDP) is shown in Figure 13. Empirical values were calculated using the World Bank data for the GDP (World Bank, 2015).

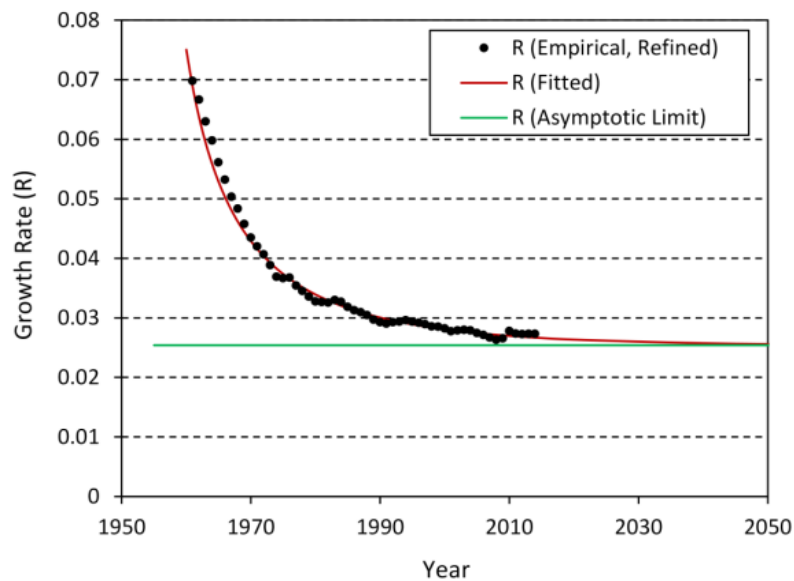

**Figure 13.** *Empirical growth rate [R (Empirical, Refined)] of the world GDP, calculated using the World Bank data (World Bank, 2015), is compared with the distribution described by the eqn (21). The growth rate approaches asymptotically a constant value. Constant growth rate generates exponential growth.*

The best fit to the growth rate data is given by the following distribution (Nielsen, 2015):

$$R \equiv \frac{1}{S}\frac{dS}{dt} = (a - be^{-rt})^{-1}. \tag{21}$$

Parameters describing empirical growth rate are: $a = 3.940 \times 10^1$, $b = 3.787 \times 10^{42}$ and $r = 4.836 \times 10^{-2}$. The asymptotic limit of the growth rate is $2.538 \times 10^{-2}$ or approximately 2.5%. The growth rate in 2014 was 2.7%, i.e. close to its asymptotic value. Constant growth rate describes exponential growth, which after a sufficiently long time is unsustainable. Even now, the growth of world GDP is approximately exponential.

Solution of the eqn (21) is given by the following distribution (Nielsen, 2015):

$$S(t) = C \exp\left[\frac{t}{a} + \frac{1}{ra}\ln(a - be^{-rt})\right]. \tag{22}$$

Its asymptotic form is represented by the exponential function,

$$S(t \to \infty) \to C \exp\left(\frac{t}{a}\right). \tag{23}$$

The distribution given by the eqn (22) is compared with the GDP data (World Bank, 2015) in Figure 14.

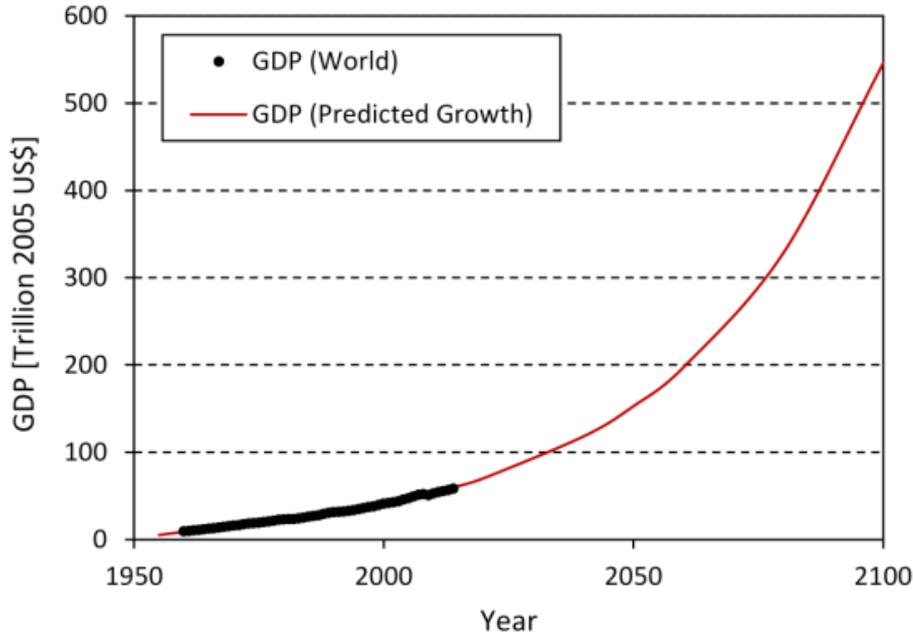

**Figure 14.** Data for the world Gross Domestic Product (GDP, World Bank, 2015), expressed in trillions of the 2005 US dollars, are compared with the distribution calculated using the eqn (22). The world economic growth is now approximately exponential and consequently, over a sufficiently long time, unsustainable.

Expressed in constant 2005 US dollars, the world GDP in 2014 was about 40% higher than in 2000. If the growth continues along the predicted trajectory, the size of the GDP in 2100 will be around 12 times higher than in 2000. Economic output of one year in 2000 will have to be generated in one month. In 2200, economic output will have to be around 170 times higher than in 2000. In one month we shall have to produce about 14 time more than we were producing in one year in the year 2000. Shall we be able to support such a dramatic increase in the economic growth? Even if we had a sufficient supply of natural resources, shall we be able to tolerate such continuing increase in the annual economic stress? Shall we be able to cope with such continuing increase of the annual economic output?

It seems to be clear that, over a sufficiently long time, exponential economic growth is unsustainable and that it will have to be changed. It is unlikely that global economic growth will be ever regulated and consequently any diversion from the exponential growth is going to happen most likely spontaneously, but spontaneous diversion is unpredictable. It might occur without any dramatic consequences but it could be also catastrophic.

A way out would be to start to decrease the growth rate along a suitably faster trajectory (if it could be done by some international agreement or by default). Such a gradual but consistently faster decrease of the growth rate could lead to a maximum in the size of the GDP or to its logistic limit (Nielsen, 2015). However, the general tendency is to increase the growth rate or at least to keep it constant. Economic status of a country is judged to be sound if its economic growth rate is high. It is therefore unlikely that the growth rate describing global economic growth will be decreasing faster than indicated by data, unless by default, which again indicates that the future of the Anthropocene is uncertain. This epoch might have a dramatic termination sooner than we expect, the termination, which could take us by surprise in much the same way as the rapid growth of population.

Maybe, with a sufficient foresight and coordinated effort we could have controlled the growth of population when its size was still small. Maybe we could have been also able to control economic growth. Malthus (1798) was wrong in claiming that the growth of population increases geometrically (exponentially). Population was never increasing exponentially but hyperbolically. However, Malthus was right when he was warning against the excessive growth of population. His warning was about 200 years ago. We had enough time to try to control the growth of population. However, his warning has been ignored in much the same way as the repeated warnings of scientists are now also consistently ignored. The Anthropocene does not have a promising future.

## 9. Summary and conclusions

The unconstrained growth of population and economic growth are hyperbolic and they were hyperbolic for the past 2,000,000 years (Nielsen, 2016d, 2016e, 2016g, 2017c). Hyperbolic growth of population was first demonstrated for the world population during the AD era (von Foerster, Mora & Amiot,1960). I have extended this early study to the BC era, first to the past 12,000 years (Nielsen, 2016e) and later to the past 2,000,000 years (Nielsen, 2017c). I have also analysed regional and global economic growth (Nielsen, 2016d) as well as regional growth of population (Nielsen, 2016g). All these studies lead to the same conclusion: the natural tendency for the growth of population and for the economic growth was to increase hyperbolically. Hyperbolic growth was so stable and so consistent that in the past 2,000,000 years there were only two major transitions between hyperbolic types of growth and only one minor disturbance of hyperbolic trajectory, which caused only small shift in hyperbolic distributions.

Hyperbolic distributions are confusing because they can be slow over a long time and fast over a short time, giving impression of being made of two distinctly different components governed by two distinctly different mechanisms of growth. These confusing features are currently erroneously interpreted, in the demographic and economic research, as representing two distinctly different stages of growth: stagnation and explosion (Nielsen, 2016a). However, hyperbolic distributions increase monotonically. There was never stagnation and never a transition to a new type of growth, which could be claimed as explosion ignited (McFalls, 2007) or detonated (Smil, 1999) by a new force of growth. The perceived explosion (the fast growth over a short time) is the natural continuation of hyperbolic growth.

The Anthropocene is characterised by the fast growth of human population and by the continually increasing, and generally devastating, impact on the environment. *The origin of the Anthropocene can be explained as the natural and even as an inevitable consequence of hyperbolic growth but the onset of the Anthropocene cannot be determined.* The eventual fast growth over a short time is the natural outcome of hyperbolic growth but there is no place on the hyperbolic distribution, which can be used to mark a transition from the slow to fast growth. There is no place, which could be identified as the beginning of the Anthropocene. This phenomenon was in the making for at least 2,000,000 years or maybe even earlier, from the dawn the existence of hominines.

The Anthropocene is not a phenomenon that emerged 100 or 200 years ago but the phenomenon that was developing over a long time. It was a slow and gradual process, which was increasing human impact on the ecology of our planet. Now, this impact is so great that it threatens even human survival. The Anthropocene was in a sense inevitable. It is a phenomenon, which is imbedded in the hyperbolic growth but it had to be hyperbolic growth of a creative species, or genus, with an endowed potential to introduce dramatic changes to the environment. These skills of using natural resources and changing the environment were also developing slowly, maybe also hyperbolically, over a long time until eventually they became dramatically manifested in changes we now observe and experience.

The Anthropocene is a feature, which represents the natural property of hyperbolic distributions, which have an intrinsic potential to increase fast over a short time, so fast that that they can increase to infinity at a fixed time. Growth to infinity is of course impossible and it has to be terminated at a certain stage. Hyperbolic distributions describing economic and population growth, global and regional, have been terminated before reaching this critical point of escape to infinity.

The major transition between 425 BC and AD 510 is the first example of such termination of hyperbolic growth of the world population close to the point of singularity, which was bypassed by only around 300 years (Nielsen, 2016e, 2017c). The currently experienced transition is another example of such a close encounter. The new singularity has been bypassed by only around 87 years, or even less if we consider that the initial minor boosting around 1950 pushed us closer to the point of singularity.

It is hardly surprising that we now have such a large size of population and that it is still continuing to increase. The initial size of the population during the AD era was much larger than the initial size during the BC era and the diversion to a new trajectory commenced much closer to the point of singularity than for the BC trajectory.

Over a long time, growth of population and the associated economic growth, were gradually and persistently following hyperbolic trajectories, which were inevitably leading to the fast growth. However, this fast growth became noticeable only in recent years and we call it now the Anthropocene. This epoch might be regarded as something new and unusual because never before had we experienced such a fast growth of human population. Never before was there also such a large number of people living on our planet.

However, even though the Anthropocene might be seen as a new phenomenon, its origin is nothing more than just the natural consequence of hyperbolic growth. Its mechanism can be also explained (Nielsen, 2016c) but its beginning cannot be determined because it is impossible to determine the time when hyperbolic growth changes from slow to fast. The change occurs gradually and monotonically over the whole range of hyperbolic distribution.

We cannot claim that the beginning of the Anthropocene was in AD 1470, the beginning of the latest hyperbolic trajectory. We cannot claim that it was in AD 510 or in 27,000 BC, the dates, which mark the beginning of the two earlier hyperbolic trajectories because all these trajectories were just the continuation of growth, which commenced much earlier, perhaps 2,000,000 years ago of perhaps even in the more remote time. This long-term growth was only relatively briefly interrupted but it was the same growth, as we can clearly see in Figure 10.

Maybe we could agree on a generally acceptable definition of the beginning of the Anthropocene but we have to understand that while its origin and mechanism can be explained, its beginning cannot be mathematically determined. Alternatively, its beginning should be traced to 2,000,000 years ago or even earlier, to the dawn of the existence of hominines.